\documentclass[twocolumn,showpacs]{revtex4}
\usepackage{times,xspace}
\usepackage{amsbsy,amssymb,amsmath,bm}
\usepackage{graphicx,color,epsfig,rotate}
\usepackage{fancyheadings}

\def\bbbc{{\mathchoice {\setbox0=\hbox{$\displaystyle\rm C$}\hbox{\hbox
to0pt{\kern0.4\wd0\vrule height0.9\ht0\hss}\box0}}
{\setbox0=\hbox{$\textstyle\rm C$}\hbox{\hbox
to0pt{\kern0.4\wd0\vrule height0.9\ht0\hss}\box0}}
{\setbox0=\hbox{$\scriptstyle\rm C$}\hbox{\hbox
to0pt{\kern0.4\wd0\vrule height0.9\ht0\hss}\box0}}
{\setbox0=\hbox{$\scriptscriptstyle\rm C$}\hbox{\hbox
to0pt{\kern0.4\wd0\vrule height0.9\ht0\hss}\box0}}}}

\newcommand{\ignore}[1]{}
\newcommand{\mComment}[1]{}
\newcommand{\gComment}[1]{}
\newcommand{\jComment}[1]{}
\newcommand{\rComment}[1]{}
\newcommand{\lComment}[1]{}

\renewcommand{\mComment}[1]{\textcolor{blue}{Manny: #1}}
\renewcommand{\gComment}[1]{\textcolor{red}{Gerardo: #1}}
\renewcommand{\jComment}[1]{\textcolor{green}{Jim: #1}}
\renewcommand{\rComment}[1]{\textcolor{magenta}{Ray: #1}}
\renewcommand{\lComment}[1]{\textcolor{purple}{Rolando: #1}}

\begin{document}
\title{Intermediate Coupling Theory of Electronic Ferroelectricity}
\author{C. D. Batista$^1$, J. E. Gubernatis$^1$, J. Bon\v ca$^2$, and H. Q. Lin$^3$}
\address{$^1$Center for Nonlinear Studies and Theoretical Division\\
Los Alamos National Laboratory, Los Alamos, NM 87545\\
$^2$ Department of Physics, FMF\\ University of Ljubljana and J. Stefan
Institute, Ljubljana, Slovenia\\
$^3$ Department of Physics, The Chinese University of Hong Kong, China}

\date{Received \today }

\begin{abstract}
We calculate the quantum phase diagram of an extended Falicov-Kimball model
for one and two-dimensional systems in the intermediate coupling regime.
Even though some features of the phase diagram are obtained analytically,
the main results are calculated with a constrained path Monte Carlo
technique. We find that this regime is dominated
by a Bose-Einstein condensation of excitons with a built-in
electric polarization. The inclusion of a finite hybridization between
the bands removes the condensate but reinforces the ferroelectricity.
\end{abstract}


\maketitle

In a recent paper \cite{batista}, one of us demonstrated that a
novel ferroelectric state, originally proposed by Portengen {\it et
al} \cite{Portengen}, is present in the strong-coupling and
mixed-valence regime of an extended Falicov-Kimball model (FKM)
\cite{Falicov}. In contrast to the traditional ferroelectric (FE)
transitions, which are a subgroup of the structural phase
transitions, the new FE state is caused by a purely
electronic mechanism of an essential {\it quantum} nature. 
Its electronic origin and the strong
interplay between the orbital and the spin flavors of the
electronic degrees of freedom make this novel state suitable for
new technological applications. For example, it opens the
possibilities of faster switching ferroelectrics and controlling
their optical properties with magnetic fields. Accordingly, it is
important to define more clearly the conditions for the
experimental observation of this novel state, especially in light
of several recent experiments that might have seen it
\cite{Moro,Spielman}.

Our extended FKM model consists of two dispersive bands 
plus an inter-band local Coulomb interaction. In absence of 
explicit hybridization between the bands, the half-filled model exhibits a
FE state that is a Bose-Einstein (BE) condensate of excitons.  The 
electric polarization is the result of a spontaneous
phase-coherent hybridization between the two local orbitals with
opposite parity under inversion. The FE moment is proportional to
the real part of the $U(1)$ order parameter of the condensate.
These conclusions were based on a strong coupling
treatment of the model. An accompanying perturbative
analysis suggested that if hybridization between the bands were
included, the BE condensate (BEC) would be replaced by a
distinguishing Ising-like FE state \cite{batista}.
Because hybridization is natural and the experimental consequences
are significant, resolving this point is important and timely.

To resolve this point, we will now step way beyond the original
work by moving away from the strong-coupling limit
and from the perturbative treatment of the hybridization 
and focus on the intermediate coupling regime.
We find that ferroelectricity remains robust.
In particular, it possesses the same phases as the strong and weak
couplings regimes, even though the nature of the condenstate changes
continuously from BCS-like in the weak coupling limit 
to a BEC of hard-core bosons in the strong couping regime \cite{Nozieres}.  
We will in fact report the complete
one (1D) and two dimensional (2D) quantum phase diagrams of the
extended FKM in the intermediate coupling regime. The results are
obtained from a scaling analysis of the numerical data produced with the Constrained 
Path Monte Carlo (CPMC) technique \cite{zhang1}. When we include a finite 
hybridization between the bands, we find as predicted by the
perturbation analysis a very large FE (or antiferroelectric)
region for the mixed valence regime 
which  is no longer a condensate of excitons but a state with an Ising-like symmetry that has 
a built-in electric polarization.  Importantly, the intermediate coupling phase diagram 
indicates more accessible regions of the ferroelectric phase than in the 
strong and weak coupling regimes. 


There are two recent experiments to which our results might apply.
Moro {\it et al} observed a FE state in free Nb clusters
\cite{Moro}. The experiment shows that cold clusters may attain an
anomalous component with very large electric dipole moments. It is
clear that in pure Nb the FE moment cannot be produced
by a lattice distortion. Therefore, the ferroelectricity must have
an {\it electronic origin}. In addition, {\it ab initio}
calculations \cite{Grad} in pure Nb indicate that the valence
electron has the same amount of $s$ and $p$ character. This
observation makes the present theory a candidate to explain
the anomalous ferroelectricity of the free Nb clusters.

Another relevant experiment is the phase coherent state
observed by Spielman {\it et al} \cite{Spielman}. The two bands are provided by
two parallel 2D electron systems interacting through an interlayer
Coulomb interaction. There is also an applied magnetic field
whose value is such that the Landau level filling factor of each layer is $1/2$.
From the results obtained in Ref. \cite{batista} and the theory that we include below,
we propose an experiment to observe electronic ferroelectriciy in these bilayer systems.

Our extended FKM for spin-less fermions on a $D$-dimensional
hypercubic lattice is:
\begin{eqnarray}
H &=& \sum_{\bf i, \nu} \epsilon_{\nu} n^{\nu}_{\bf i}
+ \sum_{{\bf i},\eta, \nu,\nu'} t_{\nu \nu'} (f^{\dagger}_{\bf i \nu} f^{\;}_{{\bf i+{\hat e}_{\eta}} \nu'}
+  f^{\dagger}_{{\bf i+{\hat e}_{\eta}} \nu'} f^{\;}_{\bf i \nu})
\nonumber \\
&+& U^{ab} \sum_{\bf i} (n^a_{\bf i}-1/2) (n^b_{\bf i}-1/2),
\label{H}
\end{eqnarray}
where $\nu=\{a,b\}$ is the orbital index, ${\bf \hat e}_{\eta}$ is
a vector in the $\eta$ direction ($\eta=\{x_1,x_2,...,x_D\}$), and
$n^{\nu}_{\bf i}= f^{\dagger}_{\bf i \nu} f^{\;}_{\bf i \nu}$ is
the occupation number of each orbital. Note that $t_{ab}=-t_{ba}$
because the two orbitals have opposite parity under spatial
inversion.

This Hamiltonian can be rewritten as an asymmetric Hubbard
model if the orbital flavor is represented by a pseudospin variable \cite{batista}:
$c^{\dagger}_{\bf i \uparrow}= f^{\dagger}_{{\bf i} a}$, $c^{\;}_{{\bf i} \uparrow} = f^{\;}_{{\bf i} a}$,
$c^{\dagger}_{\bf i \downarrow}= f^{\dagger}_{{\bf i} b}$ and
$c^{\;}_{{\bf i} \downarrow} = f^{\;}_{{\bf i} b}$.
The pseudospin components are given by:
$\tau^{\alpha}_{\bf i}= \frac{1}{2} \sum_{\nu,\nu'} f^{\dagger}_{\bf i \nu}
{\boldsymbol \sigma}^{\nu \nu'}   f^{\;}_{\bf i \nu'}$, with $\alpha=\{ x, y, z\}$.
The new expression for $H$ is an asymmetric Hubbard model with a Zeeman term:
\begin{eqnarray}
H &=& e_d \sum_{{\bf i}, \sigma} n_{i \sigma} +
\sum_{\bf \langle i, j \rangle , \sigma, \sigma'} t_{\sigma \sigma'}
(c^{\dagger}_{{\bf i} \sigma} c^{\;}_{{\bf j} \sigma'} +
c^{\dagger}_{{\bf j} \sigma'} c^{\;}_{{\bf i} \sigma})
\nonumber \\
&+& U^{ab} \sum_{\bf i} n_{{\bf i} \uparrow} n_{{\bf i} \downarrow}
+ B_z \sum_{\bf i} \tau^{z}_i,
\label{Hubbard}
\end{eqnarray}
where $e_d=\frac {1}{2} (\epsilon_a + \epsilon_b)$, $t_{\uparrow\uparrow}= t_{aa}$,
$t_{\downarrow\downarrow} = t_{bb}$, $t_{\uparrow \downarrow}=t_{ab}$,
$t_{\downarrow \uparrow}=t_{ba}$ and $B_z=\epsilon_a - \epsilon_b$.
If there is no hybridization between the bands, there is a  remaining
$U(1)$ symmetry generated by the total magnetization along the $z$ axis $M^z=\sum_{\bf i} \tau^{z}_{\bf i}$.
In the original language, this $U(1)$ symmetry corresponds to the
conservation of the difference between the total number of particles in each band
(the bands cannot interchange particles).

We will consider from now on only the half-filled case, i.e., one particle per site.
The ground state of $H$ exhibits two possible magnetic orderings
in the strong coupling limit: Ising-like antiferromagnetism and $xy$-like ferromagnetism (antiferromagnetism)
for $\gamma=-t_{\downarrow \downarrow}/t_{\uparrow \uparrow}>0$ ($\gamma<0$) \cite{batista}.
In terms of the original variables, the Ising-like phase corresponds to staggered orbital ordering (SOO)
and for $D \geq 2$, the $xy$-like ferromagnetism corresponds to a BEC of excitons with a built-in electric
polarization \cite{Portengen}:
\begin{equation}
{\bf P}= \frac {\boldsymbol{\mu}}{\Omega} \sum_{\bf i} (f^{\dagger}_{{\bf i} a} f^{\;}_{{\bf i} b}+
f^{\;}_{{\bf i} a} f^{\dagger}_{{\bf i} b})= \frac {2 \boldsymbol{\mu}}{\Omega} M^{x},
\label{pol}
\end{equation}
where $\boldsymbol{\mu}$ is the inter-band dipole matrix element and $\Omega$
is the volume of the system. The electrical polarization is then proportional
the real part of the $U(1)$ order parameter of the condensate.

An addtional particle-hole transformation on one of the spin
flavors maps Eq.~\ref{Hubbard} into a negative $U$ asymmetric Hubbard model with
the Zeeman term replaced by a chemical potential \cite{Fradkin}. Under this
transformation, the $xy$-like ferromagnetism is mapped into a
superconducting  state. This establishes a formal connection
between the BEC of excitons and a superconductor. An
eventual condensate of electron-hole pairs in the weak coupling
regime is then formally analogous to a BCS-like superconductor.

\begin{figure}[htb]
\includegraphics[angle=-90,width=6.5cm,scale=0.6]{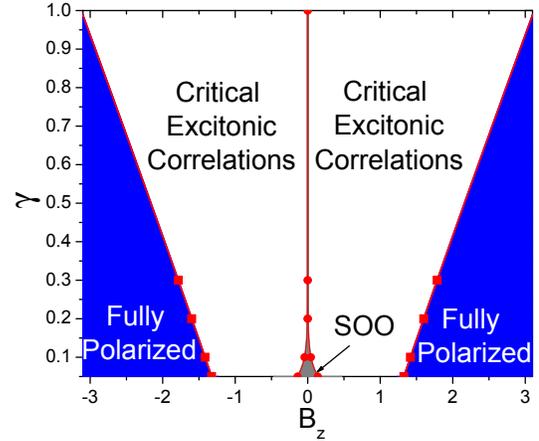}
\vspace{-0.8cm}
\caption{(color online) One dimensional quantum phase diagram of the
extended FKM $U^{ab} =1$($t_{\uparrow \downarrow}=t_{\downarrow \uparrow}=0$) \cite{note}. 
The circles are the calculated
points for $B^{c1}_z$ and the full line is an exponential fit. The error bars are smaller
than the symbol sizes. The squares are numerical results for $B^{c2}_z$ and the full
line is the analytical calculation.}
\label{fig1}
\end{figure}

{\it One dimensional case}. From now on, we will set $\gamma>0$
and $t_{\uparrow \uparrow}$ as the unit of energy. 
We start by on considering the simplest 
case of zero hybridization between the bands: $t_{\uparrow
\downarrow}=t_{\downarrow \uparrow}=0$. In Fig.~\ref{fig1}, we
present the 1D quantum phase diagram of $H$ as a function of
$\gamma$ and $B_z$. Most of the phase diagram was calculated with
the unbiased CPMC technique \cite{zhang1}. The diagram shows the same three
phases which were obtained in the strong coupling limit
\cite{batista}. For $B_z=0$, the ground state is an Ising-like
antiferromagnet (SOO). At a critical value of $|B_z|=B^{c1}_z$,
there is a transition from the orbitally ordered gaped phase to a
critical (gapless) phase with dominating pseudospin-pseudospin
correlations in the $xy$ plane. This second phase is also characterized 
by a non-zero uniform magnetization along the $z$ axis. When $|B_z|$ reaches a
second critical value, $B^{c2}_z$, $M^z$ saturates and a new gap
opens. In terms of our original language, this is the transition
from a mixed valence to non-mixed valence regime in which one band
is full and the other one is empty. This is also a transition from
an excitonic to a band insulator.

\begin{figure}[htb]
\includegraphics[angle=-90,width=7.5cm,scale=0.8]{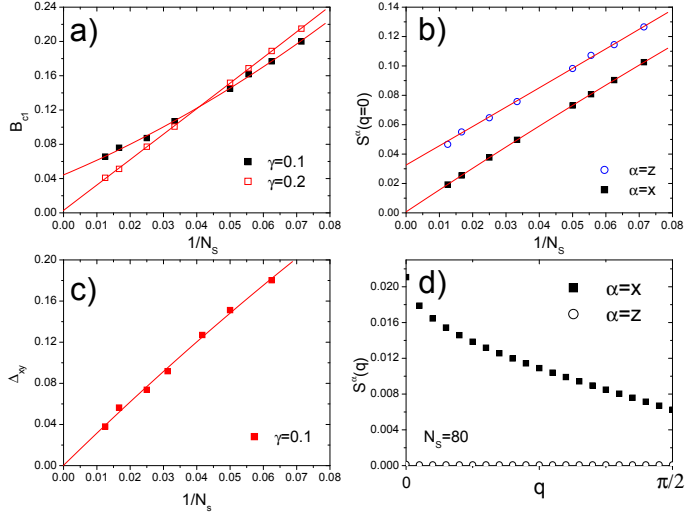}
\vspace{-1.0cm}
\caption{(color online) U=1, $t_{\uparrow\downarrow}=t_{\downarrow \uparrow}=0$. 
a) Scaling of $B^{c1}_z$ calculated with the CPMC method in chains of
length $N_s$. b) Scaling of $S^{x}(0)$ and $S^{z}(\pi)$ for $B_z=0$. 
c) Scaling of the energy gap $\Delta_{xy}$ for $M^z/N_s=1/4$. The scaling function is 
a second order polynomial expression in $1/N_s$. d) Pseudospin-pseudospin correlation 
function for $M^z/N_s=1/4$.}
\label{fig2}
\end{figure}

The value of $B^{c2}_z$ is exactly obtained by solving the one particle problem of
flipping one pseudospin in a fully saturated system. To show that this value is 
not altered by a discontinuous change of the magnetization, we also include 
the numerical calculation of $B^{c2}_z$ based on the computation of $E_{m}(M^z,B^z=0)$,
minimum energy for the subspace with a given magnetization $M^z$. Note that this can be done
because $M^z$ is a good quantum number of $H(t_{\uparrow\downarrow}=t_{\downarrow \uparrow}=0)$.
$B^{c1}_z$ is the critical value of $B^z$ which is required to change the total magnetization
$M^z$ of the ground state, i.e., it is the minimum value of $[E_{m}(M^z, B^z=0)-E_{GS}(B^z=0)]/M^z$,
where $E_{GS}$ is the ground state energy.
To calculate $B^{c1}_z$ in the thermodynamic limit, we scaled this minimum value for different chain sizes up
to $N_s=100$ ($N_s$ is the number of sites). A typical scaling is shown in Fig.~\ref{fig2}a
for two different values of $\gamma$. The finite extrapolated value of $B^{c1}_z$ implies a 
finite gap $\Delta_I$ for the Ising-like (SOO) phase. The case $\gamma=1$ corresponds to the $SU(2)$
critical point for which the pseudospin correlations along the three different axis are the same.
For the strong coupling limit \cite{batista}, it is known that the gap of the Ising-like (SOO) phase
closes exponentially (Kosterltiz-Thouless transition) when $\gamma$ gets close to one (see Fig.~\ref{fig1}).

The different phases were identified by computing the pseudospin-pseudospin correlation function:
\begin{equation}
S^{\alpha}({\bf q})= \frac{1}{N_s^2} \sum_{\bf i, j} 
(\tau^{\alpha}_{\bf i} \tau^{\alpha}_{\bf j} - \langle \tau^{\alpha}_{\bf i} \rangle 
\langle \tau^{\alpha}_{\bf j}\rangle )
e^{i{\bf q} \cdot ({\bf r_i-r_j})}
\label{str}
\end{equation}
In Fig.~\ref{fig2}b, we show the scaling of $S^x(0)$ and $S^z(\pi)$ calculated with the CPMC for
$B_z=0$ and $\gamma=0.1$. As expected, $S^z(\pi)$ has a finite value in the thermodynamic limit
indicating that the ground state has long-range staggered orbital ordering.

In Fig.~\ref{fig2}c, we show the scaling of the energy gap of the $xy$-like phase,
$\Delta_{xy}$, for $M^z/N_s=1/4$. 
$\Delta_{xy}$ scales to zero in the thermodynamic limit as expected from the critical 
nature of this phase. Fig.~\ref{fig2}d  shows that the pseudospin-pseudospin correlation 
function (see Eq.~\ref{str}) in the $x$ (or $y$) direction is much larger than the corresponding
correlation along the $z$ axis. This shows the $xy$-like nature of this phase. 

\begin{figure}[htb]
\includegraphics[angle=-90,width=6.5cm,scale=0.6]{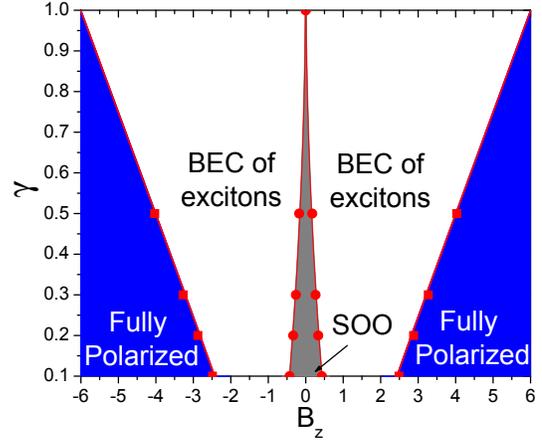}
\vspace{-0.8cm}
\caption{(color online) Two dimensional phase diagram
of the extended FKM for $U=2$
($t_{\uparrow \downarrow}=t_{\downarrow \uparrow}=0$).The circles are the calculated
points for $B^{c1}_z$ and the full line is a guide to the eye. The error bars are smaller
than the symbol sizes. The squares are numerical results for $B^{c2}_z$ and the full
line is the analytical calculation.}
\label{fig3}
\end{figure}

\begin{figure}[htb]
\includegraphics[angle=-90,width=8.0cm,scale=0.6]{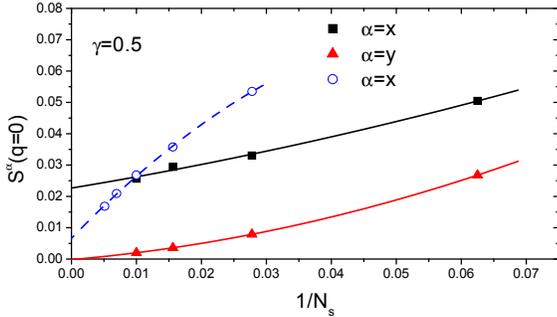}
\vspace{-4.0cm}
\caption{(color online) Scaling of $S^{x}(0)$ and $S^{y}(0)$ for a 2D system:
$U=4$,
$t_{\uparrow \downarrow}=-t_{\downarrow \uparrow}=0.2$ and $B_z=1$ (full lines),
$U=2$, $t_{\uparrow \downarrow}=-t_{\downarrow \uparrow}=0$ and $M^z/N_s=1/4$ (dashed line).
The scaling function is a second order polynomial expression in $1/N_s$.}
\label{fig4}
\end{figure}

{\it Two dimensional case}. The main qualitative difference between the 1D (Fig.\ref{fig1})
and the 2D (Fig. \ref{fig3}) quantum phase diagrams of
$H(t_{\uparrow \downarrow}=t_{\downarrow \uparrow}=0)$ is that the excitonic condensate is critical 
in one dimension, as required by the Mermin-Wagner
theorem \cite{Mermin}, and has long range order in two dimensions.
In the pseudospin language, this means that the 2D system has a finite in-plane magnetization ${\bf M}^{\perp}$ 
(see Fig.~\ref{fig4}). In addition, Eq.~(\ref{pol}) shows that this state has built-in electric polarization 
proportional to the $x$ component of ${\bf M}^{\perp}$ or the real part of the complex order parameter of the excitonic
condensate \cite{batista}.

As in the 1D case, $B^{c2}_z$($\gamma$) was computed exactly, and the
transition from the SOO to the ferroelectric BEC of
excitons is simultaneously a first order valence transition
because the relative population between the two bands, $M^z$, 
changes discontinuously. Note that in the intermediate coupling
regime, the extension of the ferroelectric phase for $\gamma$ not
very close to zero is of the order of the bandwidth while in the
strong coupling regime the size of this region is proportional to
$t_{aa}t_{bb}/U^{ab}$.

We conjecture that the three dimensional (3D) quantum phase diagram of $H$ should be similar to the
2D one since the quantum fluctuations are smaller in the first case. Moreover, we expect the 
quantum phase diagram to have the same topology for the weak and intermediate coupling regimes in any 
dimension $D \geq 2$. The main expected difference between D=2 and D=3
occurs at finite temperatures due to the Mermin-Wagner theorem \cite{Mermin}.
The transition temperature associated to the BE condensation  is finite only for the 3D case.
The BEC of electron-hole pairs undergoes a  Kosterlitz-Thouless phase transition
in a 2D system.

From Eq.~(\ref{pol}), we see that in any dimension the observed
electronic hybridization is the consequence of a spontaneous phase-coherent hybridization between the
two bands. The phase must have a real component in order to produce a finite electric polarization.
We can ask what is the physical consequence of a pure imaginary phase 
$\phi=\pm \pi/2$ ($\tan\phi=M^x/M^y$), i.e.,
a magnetization pointing in the $y$ direction. In this case, instead of ferroelectricity, the phase-coherent state
induces an ordering of atomic currents \cite{batista2}. Since the two atomic orbitals have
opposite parity, these currents do not produce a net local magnetic dipole but they can
have a net quadrupolar magnetic moment. Therefore, the $y$ direction in the pseudospin space
corresponds to ferroquadrupolar or antiferroquadrupolar magnetic ordering.

As predicted in \cite{batista}, we will now show that the inclusion of
an explicit hybridization between the bands breaks explicitly the
$U(1)$ symmetry 
which is spontaneously broken in the condensate and removes the degeneracy associated with
all the possible orientations of ${\bf M}^{\perp}$. Along with this the
natural question to answer is: what is the particular 
orientation favored by a non-zero hybridization.

{\it Inclusion of Hybridization}. In addition to the trivial $U(1)$ symmetry associated
with the conservation of the total number of particles, the remaining symmetries of $H$ for non-zero
hybridization are the spatial (${\hat I}$) and
the temporal (${\hat T}$) inversions. Since the two orbitals have opposite parities,
the signs of $t_{\uparrow \downarrow}$ and $t_{\downarrow \uparrow}$ are interchanged
under the application of ${\hat I}$. For the pseudospin components we have:
${\hat I}\tau^x_{\bf i}{\hat I}=-\tau^x_{\bf i}$, ${\hat I}\tau^y_{\bf i}{\hat I}=-\tau^y_{\bf i}$
and ${\hat I}\tau^z_{\bf i}{\hat I}=\tau^z_{\bf i}$. The temporal inversion (complex conjugation)
leaves $\tau^x_{\bf i}$ and $\tau^z_{\bf i}$ invariant and changes the sign of $\tau^y_{\bf i}$.
The ferroelectric state occurs when ${\hat I}$ is spontaneously broken. The ordering of atomic
currents occurs if both ${\hat I}$ and ${\hat T}$ are spontaneously broken. Note that the
product ${\hat I}{\hat T}$ is still a symmetry of this second state.

In Fig.~\ref{fig4}, we plot $S^x(0)$ and $S^y(0)$ as a function of the
size of a two-dimensional system.
The largest system has $N_s=100$ sites. The scaling clearly shows that the ferroelectric
state ($M_x \neq 0$) is stabilized in the thermodynamic limit. This agrees
with a perturbative analysis which is only valid in the strong coupling limit \cite{batista}.

The BEC of excitons is thus replaced by an Ising-like
ferroelectric state when a finite hybridization is included. This
result has important physical consequences since most real
systems have an explicit hybridization between two bands. This hybridization is
the reason why a BEC of excitons is so hard to observe.
The only exception is the bilayer system that we mentioned in the
introduction \cite{Spielman}. According to the theories presented
here and in Ref. \cite{batista}, this bilayer system must exhibit
the phenomenon of electronic ferroelectricity if the two layers
are made of bands with opposite parity (like "$s$" and "$p$")
instead of being identical. Then, the $d|{\bf P}|/dE$
response as a function of the uniform electric field $E$ should
show the same resonant behavior observed in the
tunneling conductance as a function of the interlayer voltage
\cite{Spielman}.

In summary, we derived the quantum 1D and 2D phase diagrams of the extended FKM
in the intermediate coupling regime. In the absence of
hybridization, we found that the insulating phase
obtained at half filling has a transition from a non-mixed valence to
a mixed valence regime as a function of the energy difference between the centers of both bands. 
At this transition the system changes from a band to an excitonic insulator which is
a BEC of electron-hole pairs. Additionally, we showed that this condensate
contains two different orderings which are degenerate: {\it ferroelectric order} and {\it chiral order}. 
The chiral state is a spontaneous ordering of local atomic currents. Since the two bands
have opposite parity under spatial inversion, these currents can only produce a non-zero quadrupolar 
magnetic moment. For bands with the same parity, the chiral state can produce orbital magnetism
and the ferroelectric state is replaced by quadrupolar electric ordering.
If the distance between both bands is further decreased, there is
a first order valence transition from the BEC of excitons to a state with
staggered orbital ordering.

Addressing an issue most important for physical systems, we also
considered the effect of a finite hybridization on these ordered
degenerate phases. We
showed that the $U(1)$ degeneracy associated with the BEC is
lifted by the hybridization, and the degenerate ground state is
replaced by an Ising-like FE state 
(broken $Z_2$ symmetry).  This result is valid even beyond the
perturbative regime $|t_{ab}|<<|t_{aa}|,|t_{bb}|$. It is crucial for
applications of the present theory to systems like the free Nb
clusters
\cite{Moro} which are most likely in the intermediate coupling
regime and have a considerable hybridization between the $s$ and
the $p$ valence bands.


We wish to thank M. P. Lilly, J. P. Eisenstein and R. Moro for stimulating discussions.
This work was sponsored by the US DOE under contract
W-7405-ENG-36. J.B. acknowledges the financial
support  of Slovene Ministry of Education, Science and Sports.
H.Q. Lin is partly supported by RGC Project CUHK 4037/02P.


\begin{references}
\bibitem{note}
The excitonic phase disappears at $\gamma=0$ due to the localization of the 
pseudospin down electrons. For D=1 at $\gamma=0$, this phase should undergo a 
transition to the most homogeneous array of down electrons or to a phase segregated
metallic state if $B_z$ is close to $B^{c2}_z$ and $U \lesssim 1.2$ \cite{far}.
A similar behavior is expected for the 2D case \cite{far}. Our results indicate 
that these transitions occur for $\gamma < 0.05$ in D=1 and $\gamma < 0.1$
in the 2D case.
\end{references}
\end{document}